\documentclass[aps,pra,twocolumn,showpacs]{revtex4-2}

\usepackage{graphicx,amsmath,amssymb}
\usepackage[toc,page]{appendix}

\begin{document}

\title{Experimental preparation of multiphoton-added coherent states of light}

\author{Jiří Fadrný}
\affiliation{Department of Optics, Faculty of Science, Palack\'y University, 17.\ listopadu 12, 77900  Olomouc, Czech Republic}

\author{ Michal Neset}
\affiliation{Department of Optics, Faculty of Science, Palack\'y University, 17.\ listopadu 12, 77900  Olomouc, Czech Republic}

\author{Martin Bielak}
\affiliation{Department of Optics, Faculty of Science, Palack\'y University, 17.\ listopadu 12, 77900  Olomouc, Czech Republic}

\author{Miroslav Ježek}
\affiliation{Department of Optics, Faculty of Science, Palack\'y University, 17.\ listopadu 12, 77900  Olomouc, Czech Republic}

\author{Jan Bílek}
\email{jan.bilek@upol.cz}
\affiliation{Department of Optics, Faculty of Science, Palack\'y University, 17.\ listopadu 12, 77900  Olomouc, Czech Republic}

\author{Jaromír Fiurášek}
\email{fiurasek@optics.upol.cz}
\affiliation{Department of Optics, Faculty of Science, Palack\'y University, 17.\ listopadu 12, 77900  Olomouc, Czech Republic}

\begin{abstract}
Conditional addition of photons represents a crucial tool for optical quantum state engineering and it forms a fundamental building block of advanced quantum photonic devices. Here we report on experimental implementation of the conditional addition of several photons. We demonstrate the addition of one, two, and three photons to input coherent states with various amplitudes. The resulting highly nonclassical photon-added states are completely characterized with time-domain homodyne tomography, and the nonclassicality of the prepared states is witnessed by the negativity of their Wigner functions. We experimentally demonstrate that the conditional addition of photons realizes approximate noiseless quantum amplification of coherent states with sufficiently large amplitude. We also investigate certification of the stellar rank of the generated multiphoton-added coherent states, which quantifies the non-Gaussian resources required for their preparation. Our results pave the way towards the experimental realization of complex optical quantum operations based on combination of multiple photon additions and subtractions.
\end{abstract}

\maketitle
 \newpage

\noindent
\section*{Introduction}

Preparation and controlled manipulation of non-classical states of light lies at the heart of quantum optics and represents a key tool for the rapidly developing optical quantum technologies. 
In optical quantum state engineering, the experimentally accessible operations include Gaussian transformations \cite{Andersen2010,Weedbrook2012} such as squeezing, interference at beam splitters, or coherent displacements, and 
single-photon detection. The photon number measurement brings in the required effective nonlinear interaction and is the key enabling tool for conditional preparation of highly non-classical quantum states of light \cite{Lvovsky2020,Biagi2022}, and for engineering transformations of optical quantum states beyond the realm of Gaussian operations. 

The most fundamental elementary quantum operations on optical modes are arguably the addition or subtraction of a single photon described by the bosonic creation and annihilation operators $\hat{a}^\dagger$ and $\hat{a}$,
respectively.  Conditional photon addition \cite{Zavatta2004,Zavatta2005PRA,Barbieri2010,Kumar20013} and subtraction \cite{Wenger2004,Ourjoumtsev2006,Nielsen2006,Wakui2007} are powerful tools in quantum photonics and they are utilized in a wide range of schemes and experiments.
Possible applications include continuous-variable entanglement distillation \cite{Opatrny2000,Ourjoumtsev2007,Takahashi2010, Kurochkin2014,Ulanov2015}, implementation of noiseless quantum amplifiers \cite{Fiurasek2009,Marek2010,Usuga2010,Zavatta2010,Park2016}, enhancement of squeezing \cite{Dirmeier2020,Grebien2022}, generation of  Schrödinger cat-like states \cite{Dakna1997,Ourjoumtsev2006,Nielsen2006,Wakui2007,Takahashi2008,Marek2008,Ourjoumtsev2009,Huang2015,Sychev2017,Asavanant2017,Takase2021,Chen2023,Endo2023}, Gottesman-Kitaev-Preskill (GKP) states \cite{Eaton2019,Konno2024} or arbitrary single-mode quantum states \cite{Dakna1999,Fiurasek2005,Nielsen2010}, preparation of hybrid entangled states of light \cite{Jeong2014,Morin2014}, and emulation of strong Kerr nonlinearity at the few-photon level \cite{Fiurasek2009,Costanzo2017}.  On the more fundamental side, coherent combinations of sequences of 
single-photon addition and subtraction enabled experimental test of the fundamental quantum commutation relation $[\hat{a},\hat{a}^\dagger]=\hat{1}$ \cite{Zavatta2009}, and the quantum-to-classical transition was studied by adding single photons to coherent states with progressively increasing amplitude \cite{Zavatta2004}.

For advanced applications and flexible quantum state engineering, simultaneous subtraction or addition of several photons is indispensable. While subtraction of up to ten photons has been demonstrated experimentally \cite{Bogdanov2017,Magana2019}, the experimental photon additions were limited to one photon \cite{Zavatta2004,Barbieri2010,Kumar20013}.
In this paper, we report on the conditional addition of up to three photons to coherent states of various amplitudes. Photons are generated in an optical parametric amplifier and coherently added to the input state directly in the nonlinear crystal. Preparation of $n$-photon-added coherent state in a signal mode of the amplifier is heralded by the detection of $n$ photons in the auxiliary idler mode. Prepared states are detected using a custom-built time-domain homodyne detector facilitating stable measurements on a time scale of hours.

We comprehensively characterize the generated quantum states by quantum state tomography. We achieve high fidelity of the generated states and observe their highly nonclassical features such as the negativity of the Wigner function. Following recent theoretical proposal \cite{Park2016}, we experimentally demonstrate that multiple photon addition enables approximate noiseless amplification of coherent states with large enough amplitude. Finally, we also analyze the certification of the stellar rank \cite{Lachman2019,Chabaud2020,Chabaud2021} of the prepared states. The stellar rank quantifies the non-Gaussian resources required for the preparation of the state. For ideal $n$-photon-added coherent states, the stellar rank is equal to the number of added photons. Our results significantly broaden the range of experimentally available elementary non-Gaussian quantum operations and pave the way towards complex engineering of quantum states and operations via combinations of multiple photon additions and subtractions.

\section*{Results}

\subsection*{\centerline{Generation of multiphoton-added coherent states}}

Photon-added coherent states (PACS) are obtained by repeated action of creation operator $\hat{a}^\dagger$ on a coherent state $|\alpha\rangle$ \cite{Agarwal1991},
\begin{equation}
|\alpha,n\rangle=\mathcal{N}_n(\alpha) \hat{a}^{\dagger n} |\alpha\rangle.
\label{PACSdefinition}
\end{equation}
Here   $\mathcal{N}_n(\alpha)=[\langle \alpha| \hat{a}^n\hat{a}^{\dagger n}|\alpha\rangle]^{-1/2}$ is a normalization factor and $n$ denotes the number of photons added to the coherent state. The $n$-photon-added coherent state  $|\alpha,n\rangle$ can be expressed as  a coherently displaced 
 superposition of the first $n+1$ Fock states \cite{Agarwal1991},
\begin{equation}
|\alpha,n\rangle=\mathcal{N}_n(\alpha)\hat{D}(\alpha)\left[ \sum_{m=0}^n \frac{n!   (\alpha^{\ast})^{n-m}}{\sqrt{m!}(n-m)!}|m\rangle\right],
\label{PACSsuperposition}
\end{equation}
where $\hat{D}(\alpha)=\exp(\alpha \hat{a}^\dagger-\alpha^\ast \hat{a})$ denotes the displacement operator.
Therefore, up to coherent displacement in the phase space, the generation of PACS $|\alpha,n\rangle$ amounts to the preparation of specific coherent superpositions of the first $n+1$ Fock states.

\begin{figure}[t!]
  \centering\includegraphics[width=0.98\linewidth]{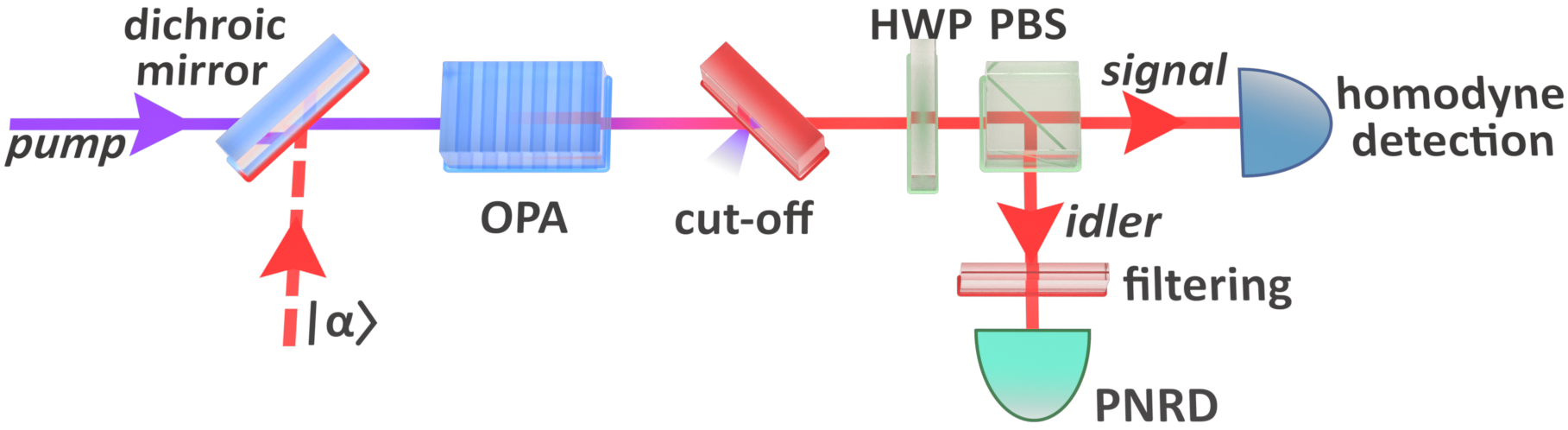}
  \caption{Schematic of the  preparation of $n$-photon-added coherent states.  A coherent state $|\alpha\rangle$ is seeded into the signal mode of an optical parametric amplifier (OPA). Once the residual pump field is filtered out, the signal and idler modes are separated by the polarizing beam splitter (PBS).  Detection of $n$ photons at the photon number resolving detector (PNRD) in the idler mode projects the signal mode to the desired  $n$-photon-added coherent state.}
  \label{fig:generation}
\end{figure}

To experimentally generate the $n$-photon-added coherent states  we utilize an optical parametric amplifier (OPA) in a pulsed single-pass regime \cite{Lvovsky2001, Zavatta2004PRA}, see Fig.~\ref{fig:generation}. The input signal mode (S) is seeded with a coherent state $|\alpha\rangle$ while the idler mode (I) is initially in a vacuum state. The nonlinear interaction in OPA generates correlated photon pairs in signal and idler modes. Detection of $n$ photons in output idler mode by a photon-number-resolving detector based 
on spatial multiplexing (see Methods) then heralds the addition of $n$ photons into the signal mode. Note that the implemented  addition of photons is probabilistic but fully heralded and does not require postselection. The generated multiphoton-added coherent states in the signal mode at the crystal output are freely propagating and are fully available and accessible for further processing and interactions. In our experiment, the generated states in signal mode are measured by a home-built balanced homodyne detector (BHD) with a 12~dB signal-to-noise ratio and 100~MHz bandwidth. More details on the experimental setup are provided in the Supplementary Information.

\subsection*{Experimental results}

In our experiment, we generate the single-photon-added coherent states,
\begin{equation}
\hat{a}^\dagger |\alpha\rangle=\hat{D}(\alpha)(|1\rangle+\alpha^\ast|0\rangle),
\label{onephoton}
\end{equation}
two-photon-added coherent states,
\begin{equation}
\hat{a}^{\dagger 2} |\alpha\rangle=\hat{D}(\alpha)(\sqrt{2}|2\rangle+2\alpha^\ast|1\rangle+\alpha^{\ast 2}|0\rangle),
\label{twophoton}
\end{equation}
and also three-photon-added coherent states
\begin{equation}
\hat{a}^{\dagger 3} |\alpha\rangle=\hat{D}(\alpha)(\sqrt{6}|3\rangle+3\sqrt{2}\alpha^\ast|2\rangle+3\alpha^{\ast 2}|1\rangle+\alpha^{\ast 3}|0\rangle).
\label{threephoton}
\end{equation}
We comprehensively characterize the generated states by homodyne tomography. We utilize the maximum likelihood algorithm \cite{Jezek2003,Hradil2004} to reconstruct the density matrix of a state from the sampled homodyne data. 
We estimate that the homodyne detection is affected by total  losses of $43\%$ (see Methods).  These losses are compensated for in the reconstruction and included in the description of the effective POVM that describes the homodyne detection.

\begin{figure*}[!t!]
  \centering\includegraphics[width=\linewidth]{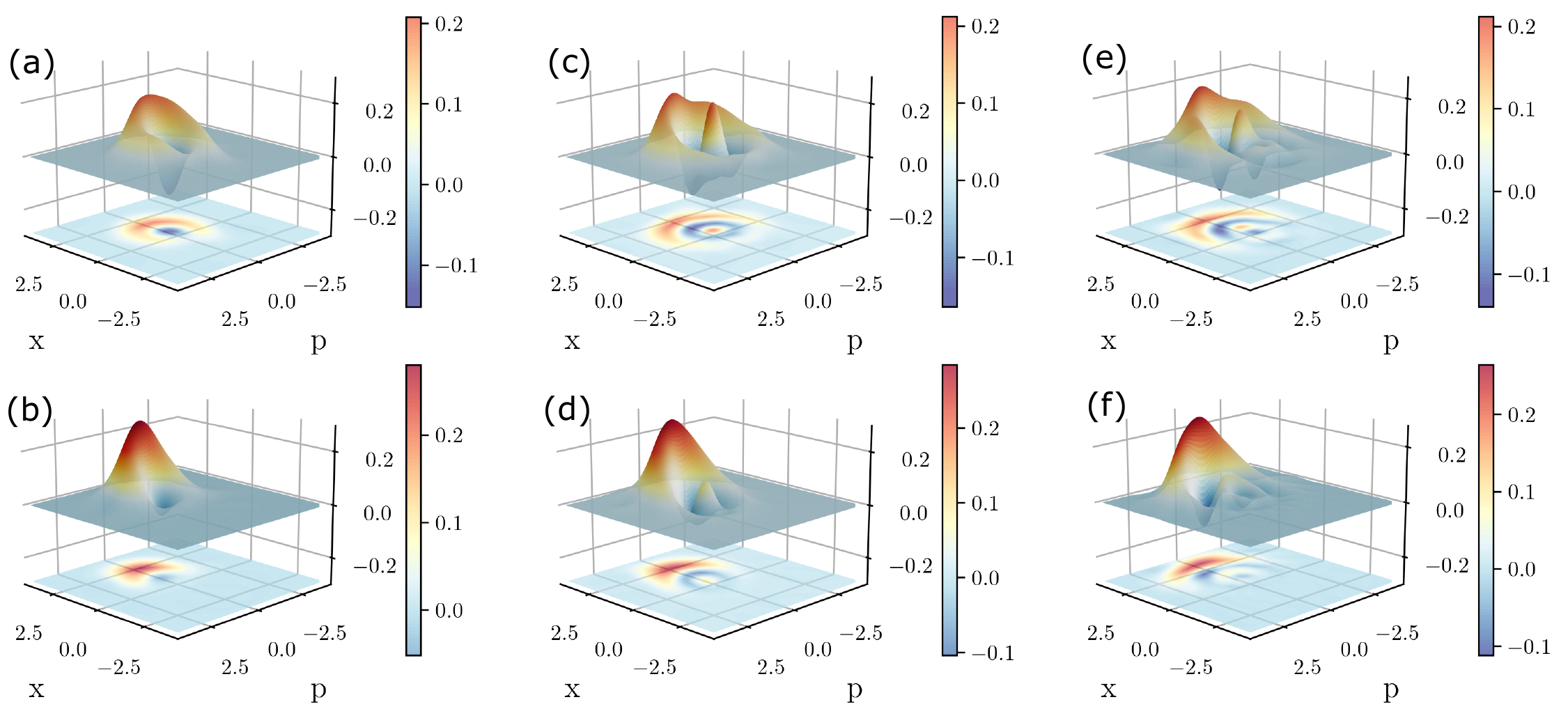}
  \caption{ Wigner functions of the experimentally generated $n$-photon-added coherent states. Shown are the Wigner functions of single-photon-added coherent states with initial seed amplitudes  $0.43$ (a) and $0.98$ (b),  Wigner functions of two-photon-added coherent states with seed amplitudes $0.34$ (c) and $0.71$ (d), and Wigner functions of three-photon-added coherent states with seed amplitudes  $0.32$ (e) and  $0.94$ (f). 
}
  \label{fig:Wigner}
\end{figure*}

\begin{table}[b!]
\caption{Fidelity $F$ and purity $\mathcal{P}$  of the experimentally generated $n$-photon-added coherent states with input amplitude $\alpha$ are listed together with the corresponding experimental probability$P_H$  of generation 
 of each state.}
\begin{ruledtabular}
\begin{tabular}{ccccc}
$n$ & $|\alpha|$ & $F$ & $\mathcal{P}$ & $P_H$   \\
\hline
1 & 0.43(1)  & 0.82(2)  & 0.72(3)& $ 8.33(4)\times10^{-5} $ \\
1 & 0.98(1) & 0.90(2)  & 0.86(4)& $ 1.84(2)\times10^{-4} $ \\
1 & 1.25(1) & 0.96(2) &  0.95(3)& $ 2.11(2)\times10^{-4} $ \\
1 & 1.43(2) & 0.97(1)  & 1.00(2)& $ 2.37(2)\times10^{-4} $ \\
1 & 1.64(1) & 0.94(2) &  0.92(5)& $ 3.03(2)\times10^{-4} $ \\
2 &  0.34(3) & 0.87(8)  & 0.9(1)& $ 9(1)\times10^{-7} $ \\
2 &  0.71(2) & 0.91(3) & 0.98(6)& $ 1.5(1)\times10^{-6} $ \\
2 &  0.96(2) & 0.94(3) & 0.97(5)& $ 2.8(2)\times10^{-6} $ \\
2 &  1.20(2) & 0.95(3) & 0.97(4)& $ 4.1(2)\times10^{-6} $ \\
2 &  1.58(3) & 0.91(3) & 0.90(5)& $ 4.7(3)\times10^{-6} $ \\
3 & 0.32(3) & 0.67(8) & 0.78(9)& $ 7(4)\times10^{-10} $ \\
3 & 0.94(3) & 0.86(6) & 0.87(8)& $ 9(4)\times10^{-10} $ \\
\end{tabular}
\end{ruledtabular}
\end{table}

 The reconstructed density matrices $\hat{\rho}$ are compared with the ideal states  (\ref{PACSdefinition}) and their fidelity $F=\langle \alpha,n|\hat{\rho}|\alpha,n\rangle$ is computed. We achieve good quality of state preparation with many of the fidelities exceeding $90\%$. 
We also calculate the purity of the reconstructed states, $\mathcal{P}=\mathrm{Tr}[\hat{\rho}^2]$, and the heralding probability $P_H$, which is determined as the ratio of the number of heralding events $N_H$ and the number of all emitted laser pulses $N_L$ within the measurement time $t$.  We summarize the parameters of the generated states in Table I. 
The amplitudes $\alpha$ in the second column of Table~I were determined from tomographic reconstructions of the input coherent states based on collected homodyne data.

In Fig. \ref{fig:Wigner} we plot Wigner functions of several experimentally generated $n$-photon-added coherent states. We can see that the generated states are highly non-classical. The experimental Wigner functions are negative in some regions of the phase space and exhibit interference patterns whose complexity increases with the number of added photons. 
All the generated $n$-photon-added coherent states also exhibit sub-Poissonian photon number distribution, as witnessed by Fano factors smaller than $1$ (see Supplementary Information).

\begin{figure*}[t]
 \centering\includegraphics[width=0.9\linewidth]{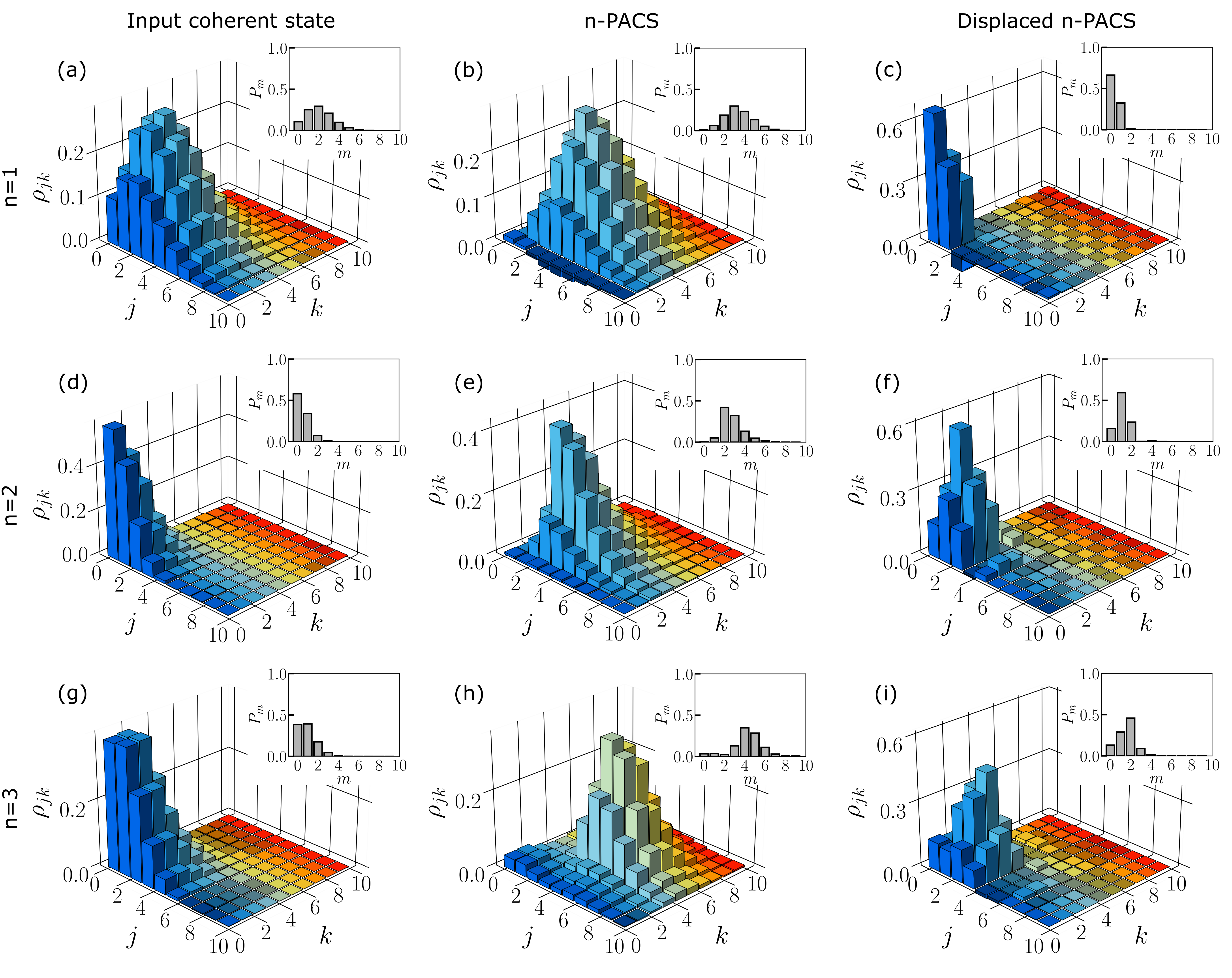}
  \caption{Examples of reconstructed density matrices in Fock basis. Density matrices of the initial coherent states are plotted in the first column. Density matrices of the experimentally generated $n$-photon-added coherent states are displayed in the second column.  Finally, the last column shows the density matrices of the $n$-photon-added coherent states coherently displaced by $\hat{D}(-\alpha)$, which in theory should result in a finite superposition of Fock states up to $|n\rangle$. The results are shown for single-photon-added coherent state with $\alpha=1.43$ (a,b,c), two-photon-added coherent state with $\alpha=0.71$ (d,e,f), and three-photon-added coherent state with $\alpha=0.94$ (g,h,i). 
Real parts of the reconstructed density matrices are plotted. The imaginary parts of the matrix elements $\rho_{jk}$ are mostly negligible and in all cases smaller than $0.075$. The insets contain plots of photon number distributions of the reconstructed states.  }
  \label{fig:rho}
\end{figure*}

Besides the Wigner functions, it is also instructive to look at the reconstructed density matrices in the Fock basis. As an example, we plot in Fig.~3  density matrices of three experimentally generated PACS. For reference and comparison, we plot in this figure also the corresponding reconstructed input coherent states. As expected, the addition of $n$ photons largely suppresses the population of the lowest Fock states 
up to $|n-1\rangle$. The remaining population of these states is caused by experimental imperfections. In order to better visualize the structure of the generated states, we apply to each reconstructed density matrix of PACS an inverse coherent displacement $\hat{D}(-\alpha_j)$, where $\alpha_j$ denotes the amplitude of the corresponding seed coherent state. The resulting states become localized in the Fock space 
and have the expected form of a superposition of the first $n+1$ Fock states (\ref{PACSsuperposition}),  and only very small fraction of the states lies outside this subspace.
The generated states exhibit similar complexity as states prepared by conditional measurement on one part of two-mode entangled state, which also enables engineering of various superpositions of Fock states \cite{Babichev2004,Ourjoumtsev2007Nature,Yukawa2013,Jeannic2018}. The advantage of our present approach is that  the heralded conditional quantum operation $\hat{a}^{\dagger n}$  can be applied to any input state, which represents an important enabling step towards  engineering of complex  non-Gaussian optical quantum operations \cite{Fiurasek2009}.

The high quality of our source of correlated photon pairs enables fast collection of sufficient data for complete tomography of the generated $n$-photon-added coherent states. Still, the heralding probability rapidly decreases with the increasing number of added photons. A single-photon addition has a typical heralding probability of $ 10^{-4}$ per pulse. 
By contrast, the typical heralding probability of two-photon addition decreases to $10^{-6}$ per pulse, and for three-photon addition it is further reduced below $ 10^{-9}$ per pulse. With the laser repetition rate of $76$~MHz, the collection of several thousands of quadrature samples requires only a few seconds for the single-PACS state. This increases to minutes for two-PACS and to several hours for the three-PACS. 
The main factor affecting the heralding probability scaling is the probability of pair generation in the nonlinear crystal. 
Nevertheless, limited single-photon detection efficiency $\eta$ and the design of our photon number resolving detector also contribute to reduced $P_H$. Just by using highly efficient superconducting single-photon detectors and a PNRD with a large enough number of channels, we could increase the heralding probability of three-PACS by an order of magnitude.

\subsection*{\centerline{Noiseless amplification of light via photon addition}}

\begin{figure}[t]
  \centering\includegraphics[width=1\linewidth]{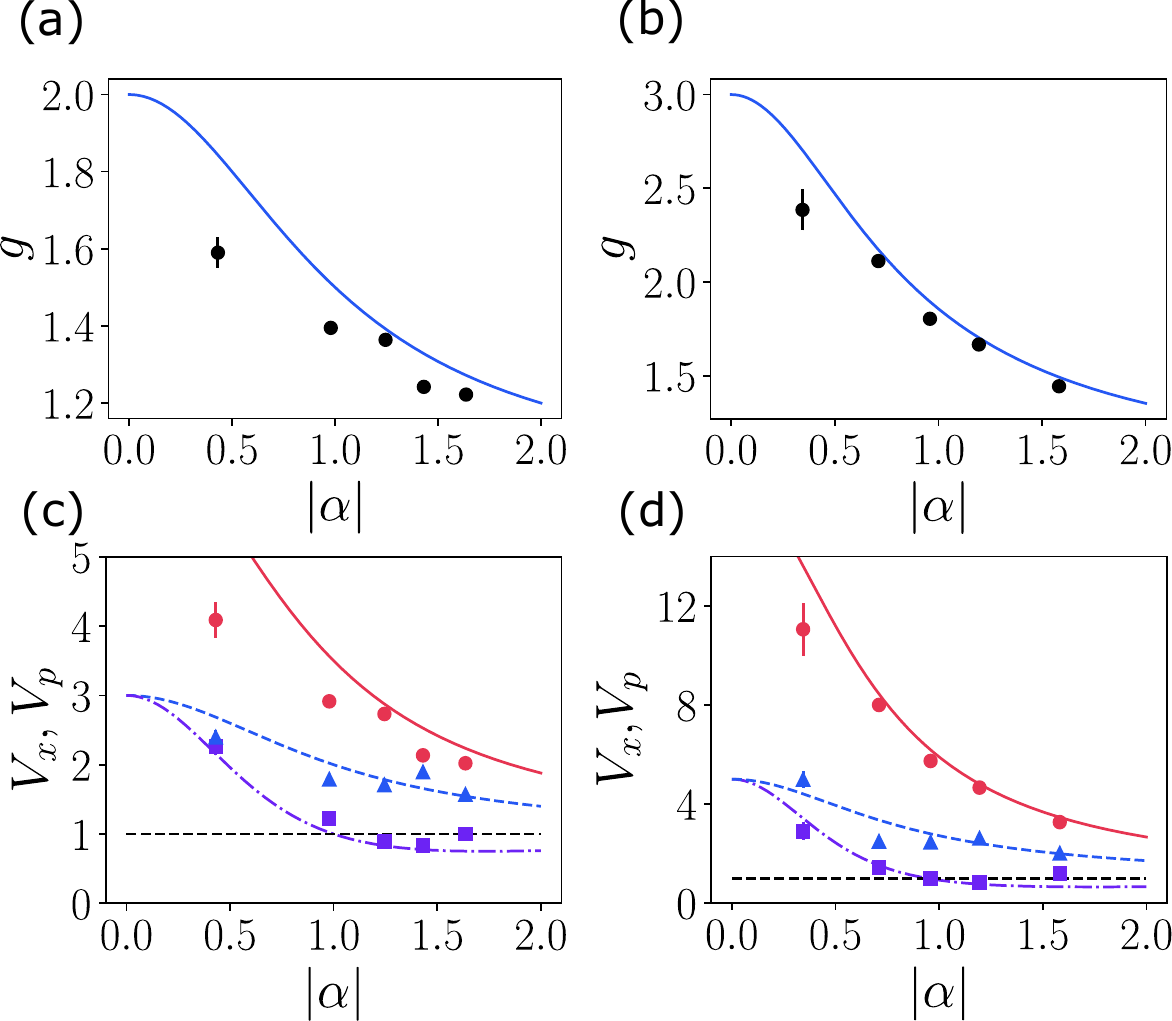}
  \caption{Characterization of noiseless amplification of coherent states by photon addition.  (a,b) The theoretical amplification gains $g_1(\alpha)$ and $g_2(\alpha)$  are plotted as lines and dots mark the experimental data.  (c,d) Experimentally determined variances of amplitude quadratures $V_x$ (purple squares) and phase quadratures $V_p$ (blue triangles) of the amplified states are plotted together with the theoretical dependencies (purple dot-dashed lines and blue dashed lines).  As a benchmark, the red solid lines in (c) and (d) indicate quadrature variance achievable by deterministic amplifier with the theoretical gains $g_1(\alpha)$ and $g_2(\alpha)$, respectively. The red dots  show variances for deterministic amplification with the 
experimentally observed gains. Vacuum variance is set to $1$ and indicated by the black dashed line. The error bars represent one standard deviation. For most of the data, the error bars are smaller than the symbol size.}
  \label{fig:gains}
\end{figure}

As theoretically shown in Ref. \cite{Park2016}, conditional photon addition can result in approximate noiseless amplification of the input coherent state (see Methods). This noiseless quantum amplifier works well for larger coherent state amplitudes, while it gives poor results for small $|\alpha|$ because it transforms a vacuum state onto Fock state $|n\rangle$.   Amplification gain of the noiseless amplifier based on conditional addition of $n$ photons can be defined as the ratio of complex amplitudes of the input and output states,
\begin{equation}
g_n(\alpha)= \frac{\langle \alpha ,n |\hat{a}| \alpha,n\rangle}{\alpha} .
\end{equation}
For $n=1$ and $n=2$  we explicitly obtain \cite{Park2016,Zhan2022}
\begin{equation}
g_1(\alpha)=\frac{2+|\alpha|^2}{1+|\alpha|^2},\qquad g_2(\alpha)=\frac{6+6|\alpha|^2+|\alpha|^4}{2+4|\alpha|^2+|\alpha|^4}.
\label{g12formula}
\end{equation}
The gain is larger than $1$, which is a signature of amplification. We plot the theoretical gains in Figs.~\ref{fig:gains}(a) and \ref{fig:gains}(b) together with the experimentally determined gains for several different values of $|\alpha|$. The gain is a decreasing function of $|\alpha|$ and it approaches $g=1$ for large $|\alpha|$. 
The experimental gains are not exactly real and exhibit small imaginary parts, caused by fluctuations and experimental imperfections. This means that the complex amplitude of the $n$-photon-added coherent state is not perfectly aligned with the amplitude of the input coherent state.  In Figs.~\ref{fig:gains}(a,b) we conservatively plot real parts of the complex gains, which for the observed small imaginary parts of the gains  is almost indistinguishable from plotting the absolute values of the gains.

As visible from the plots of Wigner functions in Fig.~\ref{fig:Wigner}, the photon-added coherent states are not fully symmetric and the variances of phase and amplitude quadratures differ. Without loss of generality, we can choose  $\alpha$ to be real in which case the amplitude and phase quadratures are defined as $\hat{x}=a+a^\dagger$ and $\hat{p}=i(a^{\dagger}-a)$, respectively. 
The quadrature variances can be expressed analytically for any $n$ \cite{Agarwal1991}, but the expressions are rather lengthy and are provided in the Supplementary Information. Here we give explicit expressions for the single-photon-added and two-photon-added coherent states. 
For $n=1$ we have

\begin{equation}
V_x=\frac{3+|\alpha|^4}{(1+|\alpha|^2)^2}, \newline \qquad V_p=\frac{3+|\alpha|^2}{1+|\alpha|^2},
\end{equation}
while for $n=2$ we obtain
\begin{equation}
\begin{split}
V_x= &\frac{20+8|\alpha|^2+12|\alpha|^4+4|\alpha|^6+|\alpha|^8}{(2+4|\alpha|^2+|\alpha|^4)^2},
\\
V_p= & \frac{10+8|\alpha|^2+|\alpha|^4}{2+4|\alpha|^2+|\alpha|^4}.
\end{split}
\end{equation}

The photon-added coherent states are non-Gaussian states, and therefore they are not minimum uncertainty states, and $V_x V_p>1$. Nevertheless,  the variances of both the amplitude and phase quadratures are smaller than the variances of the output of a deterministic linear phase insensitive amplifier with gain $g$. In this latter case
\begin{equation}
V_x=V_p=2g^2-1.
\end{equation}
In Figs. \ref{fig:gains}(c) and \ref{fig:gains}(d) we plot the theoretical dependence of quadrature variances on $|\alpha|$ together with the experimentally observed values. The quadrature variances of the experimentally generated states are indeed well below the limit of a deterministic amplifier.
This clearly demonstrates that the noise added by our probabilistic amplifier is much smaller than the noise that would be added by an ordinary linear quantum limited amplifier with the same amplification gain. This is one of the key properties and advantages of noiseless quantum amplifiers. 

The  phase quadrature variances $V_p$ are monotonically decreasing functions of $|\alpha|$. The amplitude variances $V_x$ also initially decrease but become increasing functions of $|\alpha|$ for large $|\alpha|$. Both $V_x$ and $V_p$ are maximized at $|\alpha|=0$ when the generated state is the Fock state $|n\rangle$. By contrast, we have $V_x=V_p=1$ in the limit $|\alpha|\rightarrow \infty$.  Interestingly, the amplitude quadrature $V_x$ is reduced below the level of vacuum fluctuations, $V_x<1$, for sufficiently large $|\alpha|$. Therefore, the $n$-photon-added coherent states can exhibit quadrature squeezing. This squeezing is not a consequence of interaction in a medium with high enough quadratic nonlinearity, but is a result of engineering specific superpositions of Fock states which exhibit this effect. We do observe this quadrature squeezing experimentally for specific states, see Fig.~\ref{fig:gains}.

\begin{figure}[t]
  \centering\includegraphics[width=1\linewidth]{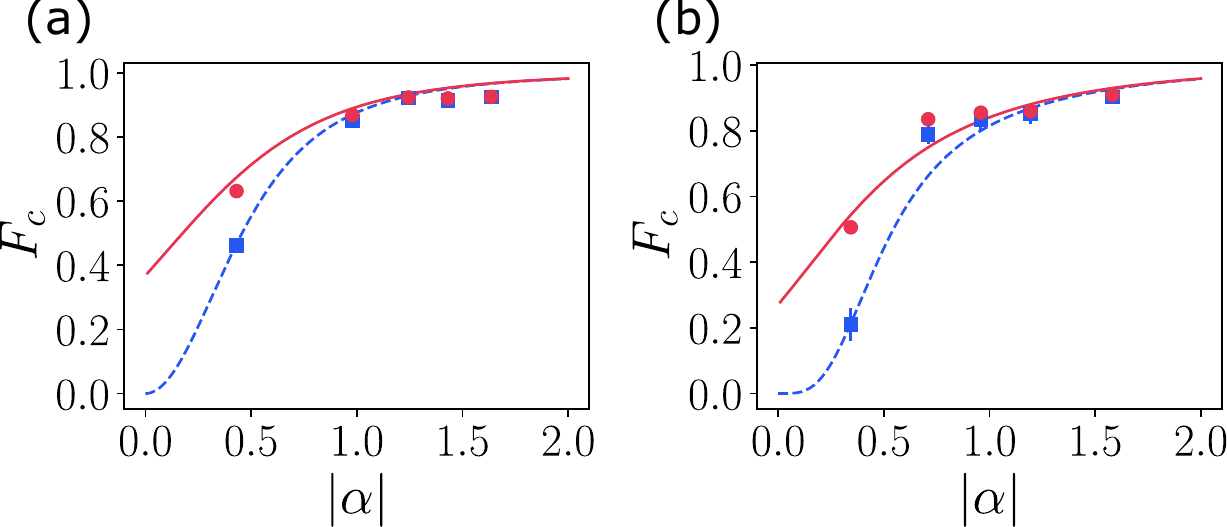}
  \caption{Fidelities of $n$-photon-added coherent states with exact coherent states. Red solid lines represent the theoretical dependence of fidelity with the coherent state with amplitude $\beta_{\mathrm{opt}}$ given by Eq. (\ref{betaopt}). Blue dashed lines depict  theoretical fidelities with the coherent states with amplitude $g_n(\alpha)\alpha$, where $g_n(\alpha)$ is the amplification gain (\ref{g12formula}). Red circles and blue squares are the corresponding experimental fidelities determined from the reconstructed density matrices of the generated states. Data are plotted for $n=1$ (a) and $n=2$ (b).
}
  \label{fig:fidelities}
\end{figure}

The closeness of the photon-added coherent state to some coherent state $|\beta\rangle$ can be quantified by fidelity $F_c(\beta)= \left|\langle \beta|\alpha,n\rangle\right|^2$.
This fidelity is maximized for 
\begin{equation}
\beta_{\mathrm{opt}}=\frac{\alpha}{2}\left(1+\sqrt{1+\frac{4n}{|\alpha|^2}}\right).
\label{betaopt}
\end{equation}
Observe that $|\beta_{\mathrm{opt}}|>|\alpha|$ yet also $|\beta_{\mathrm{opt}}|<g_n(\alpha)|\alpha|$. As illustrated in Fig.~\ref{fig:fidelities}, the maximal fidelity $F(\beta_{\mathrm{opt}})$ is a monotonically increasing function of $|\alpha|$ and asymptotically approaches unity. For comparison, we plot in Fig.~\ref{fig:fidelities} also the theoretical and experimentally determined fidelity of the $n$-photon-added coherent state with coherent state with amplitude $g_n(\alpha)\alpha$.
The graphs in Fig.~\ref{fig:fidelities} quantitatively illustrate how the similarity of the $n$-photon-added coherent state with an ordinary coherent state increases with increasing $|\alpha|$. For small $|\alpha|$ the state $|\alpha,n\rangle$ is very different from a coherent state hence the fidelity $F_c$ is small. 
In this region, the fidelity with coherent state $|\beta_{opt}\rangle$ is significantly larger than the fidelity with the coherent state $|g(\alpha)\alpha\rangle$. Consider in particular the point $\alpha =0$, where the generated $n$-photon added state is the $n$-photon Fock state $|n\rangle$. For $n>0$ fidelity of this state with vacuum vanishes, while the fidelity with a coherent state with optimal amplitude $\beta_{\mathrm{opt}}$ is nonzero. Specifically, we have $\lim_{|\alpha|\rightarrow 0}\beta_{\mathrm{opt}}=\sqrt{n}$.
It can be concluded from the fidelity plots in Fig.~\ref{fig:fidelities} that the noiseless amplification of coherent states based solely on addition of one or two photons works well  for $|\alpha|\gtrsim 1$ where the fidelities become large enough and the differences between the amplitudes $\beta_{opt}$ and $g_n(\alpha)\alpha$ practically vanish.

\subsection*{\centerline{Stellar rank of PACS}} 

The conditional photon addition generates highly non-classical and quantum non-Gaussian states. Intuitively, one expects that the nonclassicality of the state increases with the number of added photons. 
This concept can be made rigorous by introducing the so-called stellar rank of quantum states \cite{Chabaud2020}. A pure single-mode quantum state $|\psi\rangle$ is said to have a stellar rank $m$ if it can be transformed by some Gaussian unitary operation $\hat{U}_G$ to a superposition of the $m+1$ lowest Fock states from $|0\rangle$ up to $|m\rangle$,
\begin{equation}
\hat{U}_G|\psi\rangle=\sum_{n=0}^m c_n|n\rangle,
\label{UGstellar}
\end{equation}
with nonzero amplitude $c_m$. Equivalently, such states can also be called genuinely $m$-photon quantum non-Gaussian states \cite{Lachman2019}. In the experimental practice, we deal with mixed states and it is therefore important to extend the definition of stellar rank to mixed states \cite{Chabaud2020,Chabaud2021}. We say that a state $\hat{\rho}$ has a stellar rank at least $m$ if it cannot be expressed as a mixture of pure states with stellar rank $m-1$. The $n$-photon-added coherent state has stellar rank $n$, which immediately follows from Eq. (\ref{PACSsuperposition}).

\begin{figure*}[t]
  \centering\includegraphics[width=0.99\linewidth]{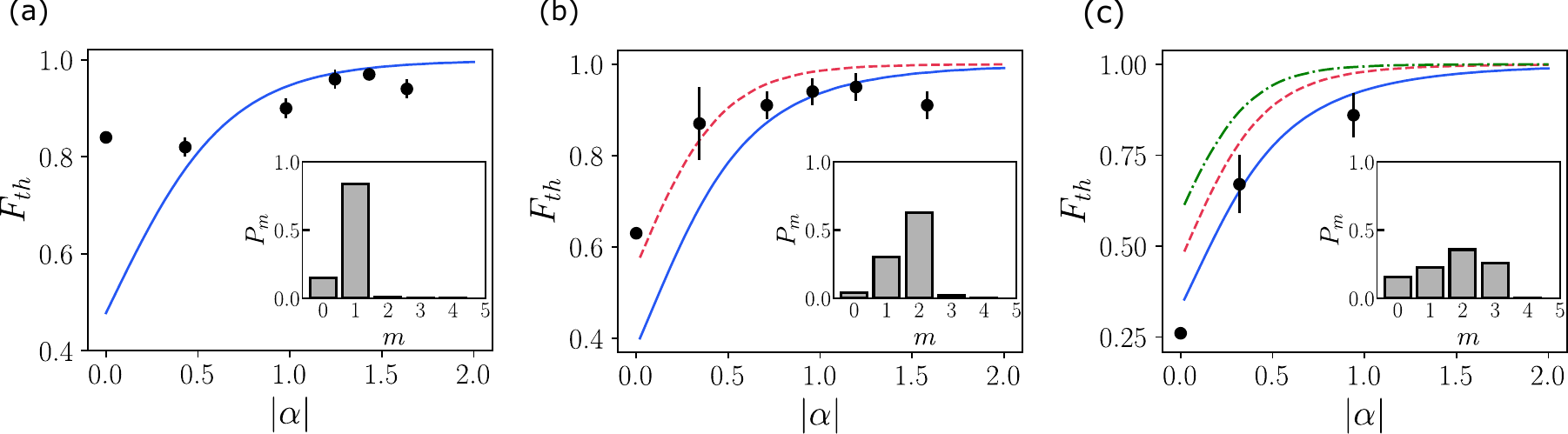}
  \caption{Fidelity thresholds $F_{\mathrm{th}}$ for certification of stellar ranks one (blue solid lines), two (red dashed lines) and three (green dot-dashed line) by fidelity with $n$-photon-added coherent state. The thresholds are plotted for $n=1$ (a),  $n=2$ (b) and $n=3$ (c). Black dots indicate the experimentally obtained fidelities, as listed in Table I. The insets show the reconstructed photon number distributions $p_n$ of the experimentally generated single-photon, two-photon, and three-photon Fock states. Fidelity of the generated Fock state $|n\rangle$ is equal to $p_n$.
} 
  \label{fig:stellar}
\end{figure*}

We can use  fidelity  $F=\langle \psi_m |\hat{\rho}|\psi_m \rangle$ of a quantum state $\hat{\rho}$ with some non-Gaussian state $|\psi_m\rangle$ with stellar rank $m$ to certify the stellar rank of $\hat{\rho}$ \cite{Chabaud2020,Chabaud2021,Fiurasek2022}. 
 For each stellar rank $k\leq m$ it is possible to establish a threshold fidelity $F_{\mathrm{th},k}$ such that the state $\hat{\rho}$ has at least stellar rank $k$  if 
\begin{equation}
F>F_{\mathrm{th},k}
\end{equation}
holds.
Let us now specifically consider the fidelity with the $n$-photon-added coherent state, which can be used to certify stellar rank up to $n$.
The corresponding fidelity thresholds for certification of various stellar ranks $k=1,2,3$ are plotted in Fig.~\ref{fig:stellar}.  Since coherent displacement is a Gaussian operation, we can equivalently consider fidelities with the finite Fock state superpositions in Eqs. (\ref{onephoton}), (\ref{twophoton}), and (\ref{threephoton}). The maximum fidelity of a coherent superposition of vacuum and single-photon state $|\psi\rangle=c_0|0\rangle+c_1|1\rangle$  with a Gaussian state was calculated numerically in Ref. \cite{Chabaud2021}. We have derived analytical formulas for the optimal Gaussian state that maximizes the fidelity, see Supplementary Information. The resulting fidelity threshold is plotted in Fig. \ref{fig:stellar}(a). The fidelity thresholds for two-photon and three-photon-added coherent states are plotted in Fig. 6(b,c) and were obtained numerically following the procedure described in Refs. \cite{Chabaud2021,Fiurasek2022}.

Figure~\ref{fig:stellar} illustrates that  the certification of stellar rank via fidelity becomes progressively more difficult with the increasing amplitude of the coherent state $|\alpha|$. This can be seen as quantification of the quantum-to-classical transition that is observed for photon-added coherent states when $|\alpha|$ increases \cite{Zavatta2004}. For large $|\alpha|$ the vacuum term in the Fock-state superpositions (\ref{onephoton}), (\ref{twophoton}), and (\ref{threephoton}) becomes dominant while for $\alpha=0$ we get the Fock state  $|n\rangle$. From the experimentally determined fidelities listed in Table I and plotted in Fig.~\ref{fig:stellar} as black dots we can certify stellar rank one for five states and stellar rank two for only one state. We note that the fidelities are not necessarily the best stellar rank witnesses and one can attempt to construct more general witnesses \cite{Fiurasek2022}. For instance, stellar rank one of the experimentally generated $n$-photon-added coherent states is witnessed by the negativity of their Wigner functions, c.f. Fig.~\ref{fig:Wigner}. 

Note also that achieving the observed fidelities of $n$-photon-added coherent states with a state of lesser stellar rank than $n$ would generally require squeezing in addition to coherent displacements and photon addition. Although squeezing is considered to be experimentally feasible Gaussian operation, implementation of a pure unitary squeezing operation on a propagating ultra-short pulsed beam of light may be challenging.  One could then modify the definition of the stellar rank and instead of all Gaussian operations $\hat{U}_G$ consider only passive single-mode Gaussian operations in Eq. (\ref{UGstellar}), i.e. coherent displacements and phase shifts. This would result in criteria with lower threshold. Investigation of such modified criteria is beyond the scope of the present paper and will be the subject of future work.

Figure~\ref{fig:stellar} suggests that certification of high stellar rank may be easiest for $\alpha=0$, when Fock state $|n\rangle$ is generated under ideal conditions. We plot in the insets of Fig.~\ref{fig:stellar} the photon number distributions of the experimentally generated Fock states $|1\rangle$, $|2\rangle$ and $|3\rangle$, respectively. Indeed, the fidelities of Fock states $|1\rangle$ and $|2\rangle$ are large enough to certify the stellar rank one and two, respectively. On the other hand, the data for Fock state $|3\rangle$ are deteriorated due to low generation rate and long acquisition time, so than only stellar rank one can be certified for this state. Especially for Fock states, the certification of their stellar rank from homodyne detection appears to be more difficult than certification based on photon counting measurements \cite{Lachman2019}. Certification of stellar rank of Fock states
based on photon counting measurement can be achieve by observation of absence (or great suppression) of the $n+1$-fold coincidences with respect to the $n$-fold coincidences. On the other hand, homodyne detection probes the photon number statistics only indirectly, and the state must be reconstructed by quantum tomography, which results in nonzero (albeit small) tail in the reconstructed photon number distribution. Nevertheless, the coherent homodyne detection is absolutely crucial in our experiment, because it allows full characterization of the generated states, including their coherences, off-diagonal density matrix elements in Fock basis, and complete Wigner functions.

\section*{Discussion}
In summary, we have successfully experimentally implemented conditional addition of up to three photons to coherent states of light. The input coherent state  is injected into the signal mode of an optical parametric amplifier and successful addition of $n$ photons is heralded by detection of $n$ photons in the output idler mode of the amplifier.  We have completely characterized the generated state by a home-built time-domain homodyne detector with stable long-term operation over the time of 4 hours. The experimentally generated $n$-photon-added coherent states exhibit high fidelities, negative Wigner functions with complex interference patterns in phase space, and sub-Poissonian photon number statistics. 
Our analysis of the stellar rank of the generated photon-added coherent states confirms that our experimental setup provides an advanced non-Gaussian quantum resource. As an application, we have experimentally demonstrated  approximate noiseless quantum  amplification of coherent states by conditional photon addition, which works well for coherent states with not too small amplitude. 
The demonstrated results significantly broaden the range of experimentally available elementary non-Gaussian quantum operations and pave the way towards the experimental realization of complex optical quantum operations based on  combination of multiple photon additions and subtractions. 

\begin{figure}
\centerline{\includegraphics[width=0.8\linewidth]{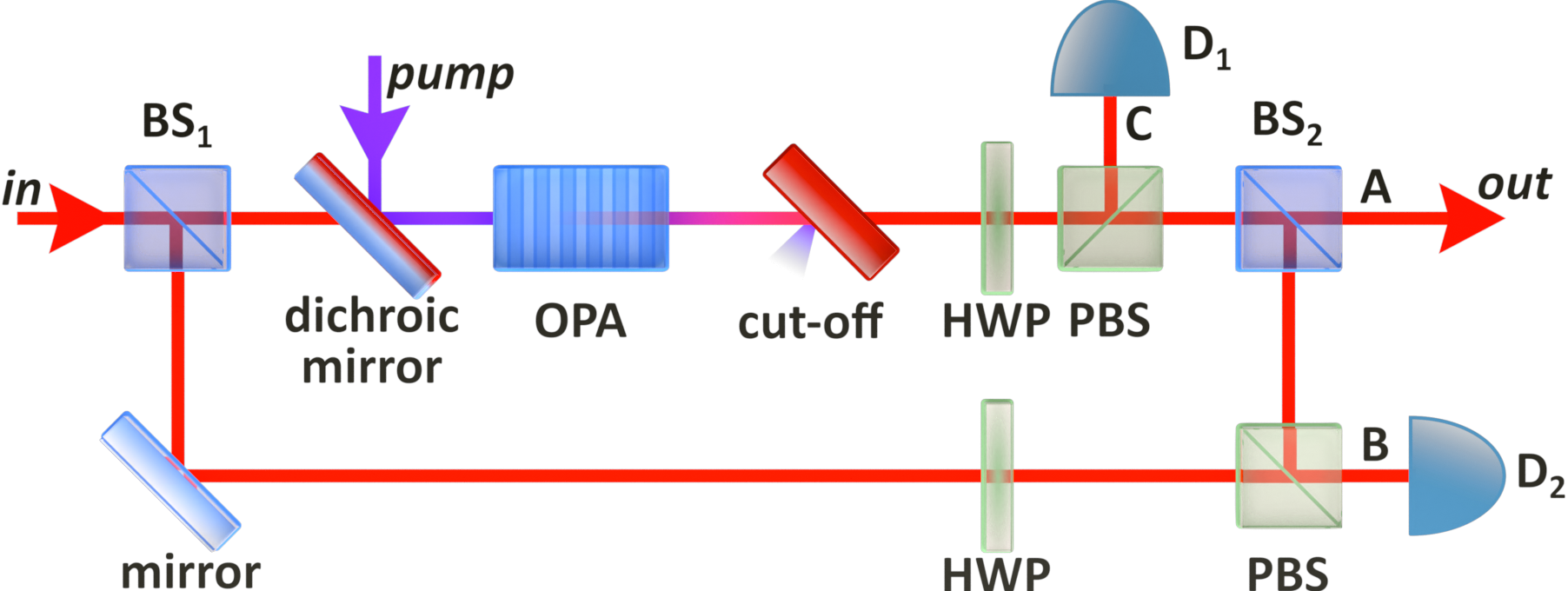}}
\caption{Schematic of implementation of coherent superpositions of various sequences of additions and subtractions of $N$ photons \cite{Fiurasek2009,Zavatta2009,Costanzo2017}. Detection of $N$ photons at the photon number resolving detector D$_1$ heralds the  addition of $N$ photons. Photons can be subtracted both before and after the addition using two unbalanced beam splitters BS$_1$ and BS$_2$. Detector D$_2$ projects the two polarization modes onto  a specific pure $N$-photon polarization state which erases the which-way information and creates coherent superposition of various sequences of $\hat{a}^{N-j} \hat{a}^{\dagger N} \hat{a}^{j}$.}
\label{fig:polynomial}
\end{figure}

In particular, by realizing coherent combinations of various sequences of additions and subtractions of $N$ photons, we can implement arbitrary operations diagonal in Fock basis which can be expressed as polynomials of $N$-th order in the photon number operator $\hat{n}$ \cite{Fiurasek2009}. This would extend the already experimentally demonstrated coherent superpositions of single-photon additions and subtractions $\hat{a} \hat{a}^\dagger$ and $\hat{a}^\dagger \hat{a}$ \cite{Zavatta2009,Costanzo2017} to higher $N$.  A possible implementation is illustrated in Fig.~\ref{fig:polynomial}. The photons can be subtracted both before and after the photon addition \cite{Fiurasek2009,Zavatta2009,Costanzo2017}. The two beams that may contain the subtracted photons are spatially recombined at polarizing beam splitter PBS and projected onto a specific pure $N$-photon polarization state. This projection, together with the heralded addition of $N$ photons, creates the coherent superposition of various sequences $\hat{V}_j=\hat{a}^{N-j} \hat{a}^{\dagger N} \hat{a}^{j}$, with $0 \leq j\leq N$. Eech term $\hat{V}_j$ is a polynomial of $N$-th order in photon number operator. By projecting onto various $N$-photon polarization states we can generate  arbitrary linear combinations of the $N+1$ independent polynomials $\hat{V}_j$. Additional details are provided in the Supplementary Information, including an example of implementation of the single-mode nonlinear sign gate relevant in the context of quantum computing with linear optics.
 
The achievable success probability of multiphoton addition can be significantly increased by addressing the limitations of our present setup.  First, the limited detector efficiency of the utilized avalanche photodiodes and the design of our photon number resolving detector contribute to reduced 
 success probability. Second, we have observed a low damage threshold of our PPKTP nonlinear crystals used in the OPA which strongly limited the gain of the OPA and forced us to operate it with a relatively weak pump beam focusing resulting in lower brightness of the source. By using more robust nonlinear crystals, highly efficient superconducting single-photon detectors, and increasing the number of detection channels of the multiplexed PNRD, the success probabilities can be increased by several orders of magnitude thereby paving way to  experimental addition of even higher number of photons. Already with the present setup the experimental combination of conditional addition of two photons and conditional subtraction of two photons appears to be feasible and will be the subject of future work. 

\bigskip

\section*{Methods}

\subsection*{\centerline{PNRD based on spatial multiplexing}}
Our PNRD is based on spatial multiplexing and consists of a combination of two tunable optical beam splitters formed by a half-wave plate and polarizing beam splitter, and three single-photon avalanche diodes serving as single-photon detectors (SPD) that can distinguish the presence or absence of photons. The addition of $n$ photons is heralded by $n$-fold coincidence click of the detectors. We use a specific beam division configuration for each $n$. For single-photon addition, we send the whole idler beam to $\mathrm{SPD}_{1}$. If we want to herald a two-photon addition, we evenly split the idler mode to $\mathrm{SPD}_{1}$ and $\mathrm{SPD}_{2}$. Finally, to herald the three-photon addition we evenly split the idler mode among all three detectors.  Electronic outputs from SPDs are processed in a custom programmable coincidence unit. The coincidence unit allows to set coincidence window in the range of 0.5-5 ns, while a tunable time delay  for each input channel is used for synchronization. 

In the spatially multiplexed PNRD, multiple photons can sometimes impinge on the same SPD, which reduces the success probability of multiphoton detection. 
With our PNRD design, we can detect the incoming single photon with probability $\eta$, two photons with probability $\eta^2/2$ and three photons with probability $2\eta^3/9$, where $\eta$ is the detection efficiency of the SPD. The heralding can be also affected by false triggers arising from higher pair events, when at least $n+1$ photons are present in the idler beam but only $n$-fold coincidence detection is observed. However, since the probability of pair generation in the crystal is less than $0.01$ these higher photon contributions are negligible. 

In our experiment, we change the configuration of the heralding photon-number resolving measurement in dependence on the targeted number of added photons $n$. However, this is done for technical reasons only. Specifically, the  utilized coincidence logic did not allow us to evaluate more than one coincidence pattern. With a more sophisticated coincidence unit 
it would be possible to use a  single fixed conditioning measurement scheme with several channels and several single-photon detectors, and process the signal from the detectors to count the number of detected idler photons at each run of the experiment. The heralding signal would then provide the information how many photons were added to the input state. In typical applications of  photon addition one  usually aims to perform some specific quantum operation which  involves a well defined fixed number of photon additions. Therefore,  the need to reconfigure the conditioning device depending on the number of targeted photon additions is not a significant limitation.

\subsection*{\centerline{Estimation of the effective efficiency of homodyne detection}}
We estimate that the homodyne detection is affected by total  losses of $43\%$. This includes losses due to the filtering optical fiber inserted in the path of the output signal beam ($T_1=0.80$), losses in other optical components after the crystal output ($T_2=0.89$),  limited visibility of interference with the local oscillator ($V=0.96$, $T_3=0.92$), quantum efficiency of the photodiodes in homodyne detector ($T_4=0.92$) and effective losses caused by finite SNR of the homodyne detector ($T_5=0.944$). The losses can be estimated as $1-T$, where $T=\prod_{j=1}^5 T_j$ is the overall effective transmittance. 
Our calibration of the homodyne detector efficiency and the resulting estimation of the coherent-state amplitudes $\alpha$ can be cross-checked by measurements of the relative heralding probabilities of preparation of the $n$-photon-added coherent states. These measurements, which do not involve data from homodyne detector, confirm the consistency of our estimation of $\alpha$ (see Supplementary Information).

The optical fiber inserted in the path of the output signal beam is not strictly necessary, although there is some convenience in having the fiber in the setup.  The alignment of the signal with the homodyne local oscillator is simplified because the signal and the local oscillator are outcoupled by the same type of outcoupler, ensuring good mode matching and thus visibility of the homodyne detection. Also, fiber-based distribution of the signal  is space efficient. If we do not include the losses imposed by the fiber into the total detection efficiency budget, we still obtain  losses of $29\%$, given by $1- \prod_{j=2}^5 T_j$. For  comparison, we have performed reconstruction of the generated quantum states without compensating for the losses imposed by the fiber. The fidelities and purities of the reconstructed states remain high and are typically reduced only by a few percent with respect to the full loss compensation. For details, see Table S1 in the  Supplementary Information.

\subsection*{\centerline{Noiseless amplification via photon addition}}
Deterministic noiseless amplification of coherent states is forbidden by the laws of quantum mechanics.  An ideal probabilistic noiseless quantum amplifier is described by the operator $g^{\hat{n}}$, with $\hat{n}=\hat{a}^\dagger\hat{a}$ and $g>1$, and we have \cite{Ralph2009}
\begin{equation}
g^{\hat{n}}|\alpha\rangle=e^{(g^2-1)|\alpha|^2/2}|g\alpha\rangle.
\label{noiselessideal}
\end{equation}
The operator $g^{\hat{n}}$ is unbounded and therefore unphysical.  Experimental implementations of the noiseless quantum amplifiers \cite{Zavatta2010,Usuga2010,Xiang2010,Ferreyrol2010,Osorio2012,Kocsis2013} therefore approximate the unphysical operation (\ref{noiselessideal}). Noiseless amplification requires modulation of amplitudes of Fock states such that the amplitudes of higher Fock states are enhanced with respect to the amplitudes of the low Fock states.
A canonical example is the noiseless quantum amplifier based on the combination of conditional photon addition and subtraction, which applies the operation $\hat{a}\hat{a}^\dagger =\hat{n}+1$ to the input state \cite{Fiurasek2009,Marek2010,Zavatta2010}. 
Alternative implementations based on quantum scissors enhance the amplitude of the single-photon state with respect to the amplitude of the vacuum state, while removing the rest of the state \cite{Ferreyrol2010,Xiang2010}. One may then wonder how the noiseless amplification based solely on photon addition fits into this picture. 

To get some insight, recall that coherent state is the eigenstate of the annihilation operator, $\hat{a}|\alpha\rangle=\alpha|\alpha \rangle$. Therefore, the state $\hat{a}^{\dagger n}|\alpha\rangle$ is identical to the state $\hat{a}^{\dagger n} \hat{a}^n| \alpha\rangle$, up to a normalization factor. Thus, for input coherent states, the addition of $n$ photons is equivalent to the application of a quantum filter diagonal in Fock basis, as far as the generated state is concerned. The difference is only in the success probability of the applied conditional operation. In particular, for $n=1$ we get the Fock diagonal filter $\hat{n}$, while for $n=2$ we get $\hat{n}(\hat{n}-1)$. For not too small complex amplitudes $\alpha$, these Fock-amplitude modulations reasonably well approximate the noiseless amplification (\ref{noiselessideal}). 

In the experiment, we reconstruct both the photon-added coherent state and the corresponding input coherent state and determine the effective noiseless amplification gain as the ratio of their complex amplitudes. We emphasize that the determination of the gain is not affected by the efficiency $\eta_H$ of the homodyne detector, because this factor cancels out when we calculate the ratio of two complex amplitudes, 
\begin{equation}
g_{\mathrm{exp}}=\frac{\alpha_{\mathrm{out}}}{\alpha}=\frac{\eta_H \alpha_{\mathrm{out}}}{\eta_H \alpha}.
\end{equation}

\section*{Data availability}
Data is available from the corresponding authors upon reasonable request.

\section*{Code availability}
The codes used to generate data for this paper are available from the corresponding authors upon reasonable request.

\section*{Acknowledgements}
We acknowledge the financial support of the Czech Science Foundation (Project No. 21-23120S) and the Palacký University under Projects No. IGA-PrF-2023-002 and IGA-PrF-2023-006. J. Fadrn\'{y} and J. Fiur\'{a}\v{s}ek
acknowledge the project 8C22002 (CVSTAR) of MEYS of the Czech Republic, which has received funding from the European Union’s Horizon 2020 Research and Innovation Programme under Grant Agreement no. 731473 and 101017733.

\section*{Competing interests}
All authors declare no financial or non-financial competing interests.

\section*{Author contributions}
J.~Fadrn\'{y} and MN performed the experiment, data processing, and numerical simulations. MB participated in the experiment and the source development.
MJ initiated the experimental project and participated in the experiment, data processing, and interpretation. JB coordinated the experimental project and participated in the experiment, data processing, and interpretation. JB a MJ developed the time-domain homodyne detection. J.~Fiur\'{a}\v{s}ek coordinated the project, developed the theory, and participated in data interpretation.
J.~ Fiur\'{a}\v{s}ek and J.~Fadrn\'{y} wrote the manuscript, and all authors were involved in revising the manuscript.

\section*{Corresponding authors}
Correspondence to Jan B\'{\i}lek or Jarom\'{i}r Fiur\'{a}\v{s}ek.

\clearpage
\onecolumngrid

\begin{center}
\textbf{\large Supplemental Materials}
\end{center}

\setcounter{equation}{0}
\setcounter{figure}{0}
\setcounter{table}{0}
\setcounter{page}{1}

\makeatletter
\renewcommand{\theequation}{S\arabic{equation}}
\renewcommand{\thefigure}{S\arabic{figure}}
\renewcommand{\bibnumfmt}[1]{[S#1]}
\renewcommand{\citenumfont}[1]{S#1}

\section*{Detailed description of the experimental setup}

\begin{figure*}[b!]
  \centering\includegraphics[width=\linewidth]{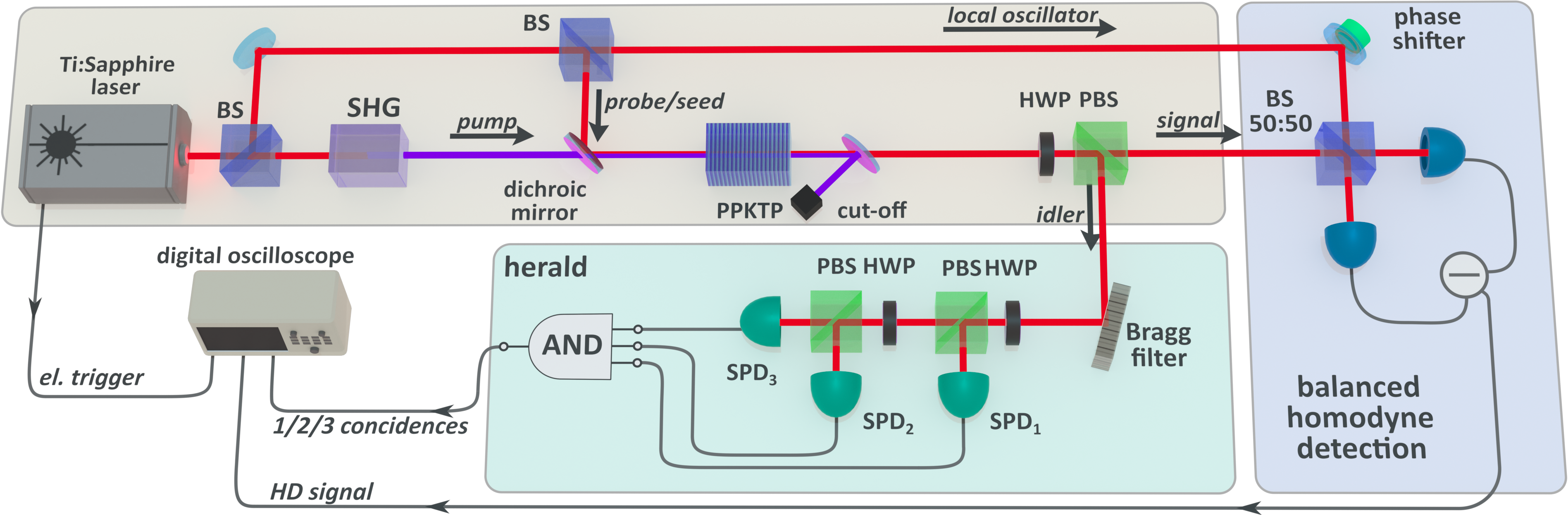}
  \caption{Experimental setup. BS, beam splitter; SHG, second-harmonic generation; PPKTP, periodically polled potassium titanyl phosphate crystal, HWP, half-wave plate; PBS, polarizing beam splitter; SPD, single-photon detector; AND, coincidence logic gate.}
  \label{fig:setup}
\end{figure*}

A schematic of of the full experimental setup is depicted in Fig.~\ref{fig:setup}. The optical power is provided by a pulsed Ti: Sapphire laser (Coherent Mira HP) outputting 1.5 ps pulses at a central wavelength of 800 nm with a 76 MHz repetition rate. This primary laser beam is split into three main paths - the fundamental beam for second harmonic  generation, a local oscillator for homodyne detection, and a seed/probe beam for stimulated parametric down-conversion and alignment purposes.

Most of the laser power is converted to a second harmonic at 400 nm in the SHG module. The second harmonic pulses pump a periodically poled potassium titanyl phosphate crystal (PPKTP) where correlated photon pairs are generated in the process of parametric frequency down-conversion.  The crystal is 2 mm long with a 9.25 $\mu$m poling period.  The pump beam is focused to the middle of the crystal with a beam waist of 33 $\mu$m. Collinear degenerate type-II phase matching is achieved by temperature tuning of the crystal. 
The residual part of the pump at the output of the nonlinear crystal is filtered out by a cut-off filter. Due to type-II phase matching the generated photon pairs are orthogonally polarized and they are spatially separated by a polarizing beam splitter PBS to signal and idler modes. The idler mode is first spatially filtered by coupling into a single-mode optical fiber and then undergoes narrow spectral filtering using a Bragg filter with a FWHM of 0.4 nm.  This ensures that the photon-added coherent states are conditionally prepared in a well defined single spatiotemporal mode and have high purity.  A small part of the fundamental beam serves as the initial coherent state for the photon addition. It is mode-matched to the signal mode in the PPKTP crystal  using a system of lenses and mirrors, thus seeding the parametric process. Time alignment of pump and seed pulses is ensured by a delay line in the path of the seed beam.

The coherent nature of the photon addition demands high purity of the source and perfect mode matching in the nonlinear crystal. The utilized  single-pass source operated in the pulsed regime is notorious for deteriorating effects such as gain-induced diffraction and dispersion effects. Dozens of experimental and material parameters play a role in the process. For example, stronger focusation in the nonlinear crystal and shorter pulses make the deteriorating effects more pronounced. Weaker focusation and long pulses might remedy these issues but make the generation rate very low, which represents a major roadblock to perform the experiment. We developed our  source to operate close to the optimum configuration within the mentioned trade-offs. We increase the pump power when targeting  higher-photon addition to increase its  success probability.  The peak density power of the pump is so high during the $3$-photon adding measurements that the crystal must be continuously shifted  to prevent local damage to the crystal.

We can prepare single-, two-, or three-PACS in the signal mode by heralding on one, two, or three photon detection at the photon number resolving detector (PNDR) in the idler mode. 
The generated states in the signal mode are coupled in a single-mode optical fiber and then coupled out and combined on a 50:50 beam splitter (BS) with the local oscillator (LO). 
The fiber coupling allows for distributing the prepared states to distant parts of the experiment and thus enhances the flexibility of the setup.
Pulse synchronization at the BS is achieved by a delay line in the path of the local oscillator, which includes a piezo-driven mirror stage for the precision phase scanning. 

The states are then measured by the balanced homodyne detector (BHD) with $92\%$ quantum detection efficiency, 12~dB signal-to-noise ratio and 100~MHz bandwidth. The electronic output of the BHD is sent to a fast digital oscilloscope together with the PNDR heralding output and electronic trigger from the laser. The homodyne detection is heralded by up to three single-photon detections in the heralding arm and the master clock of the laser at 76 MHz. This was achieved by a home-built high-throughput coincidence logic with a coincidence window as short as 0.5 ns and approximately 10 ps jitter with respect to the master clock.

Each coincidence detection triggers the acquisition of the homodyne signal to determine the quadrature value of the measured $n$-PACS and the phase of the local oscillator at the time of the coincidence. The quadrature value is extracted directly from the homodyne signal at the detection time. To determine the phase, the time period of around 250~$\mu$s after the coincidence detection is sampled by 100 quadrature measurements. These quadratures are unheralded, thus depicting the phase evolution of the seed coherent state and allowing us to estimate the phase of the local oscillator with respect to the coherent state.

\section*{Theoretical model of single-pass optical parametric amplification}
The action of the single-pass OPA is described by a unitary operation
\begin{equation}
\hat{U}_{\mathrm{OPA}}=\exp\left(\kappa\hat{a}^{\dagger}_{\textrm{S}} \hat{a}^{\dagger}_{\textrm{I}} -\kappa \hat{a}_{\textrm{S}} \hat{a}_{\textrm{I}} \right)
\end{equation}
where $\kappa$ is an effective interaction strength. This unitary operation can be decomposed as follows,
\begin{equation}
U_{\mathrm{OPA}}=e^{\lambda \hat{a}_S^\dagger \hat{a}_I^\dagger}(1-\lambda^2)^{(\hat{a}_S^\dagger \hat{a}_S+\hat{a}_I^\dagger \hat{a}_I+1)/2} e^{-\lambda \hat{a}_S\hat{a}_I},
\label{UOPAdecomposition}
\end{equation}
where $\lambda=\tanh(\kappa)$. For the input state $|\alpha\rangle_S|0\rangle_I$, the output state of signal mode conditional on detection of $n$ photons in the idler mode reads
\begin{equation}
|\psi\rangle=\sqrt{\frac{1-\lambda^2}{n!}} \lambda^n  e^{-\lambda^2|\alpha|^2/2} \hat{a}_S^{\dagger n} |\sqrt{1-\lambda^2}\alpha\rangle,
\end{equation}
The addition of $n$ photons is accompanied by a noiseless attenuation \cite{Micuda2012,Nunn2021,Shringarpure2022} of the input state by factor $\sqrt{1-\lambda^2}$, c.f. the term $(1-\lambda^2)^{\hat{a}^\dagger_S\hat{a}_S/2}$ in Eq. (\ref{UOPAdecomposition}). Since our experiment is operated in the regime $\lambda \ll 1$, this noiseless attenuation effect can be neglected. The probability of detection of $n$ photons in the idler mode reads
\begin{equation}
P_H(\alpha,n)=(1-\lambda^2) \lambda^{2n}L_n\left(-(1-\lambda^2)|\alpha|^2\right)e^{-\lambda^2|\alpha|^2},
\label{PH}
\end{equation}
where   $L_n(x)$ denotes a Laguerre polynomial of degree $n$.
Multiphoton addition requires sufficiently strong interaction in the OPA because the heralding probability scales as $\lambda^{2n}$.

\begin{figure}[t]
  \centering\includegraphics[width=0.75\linewidth]{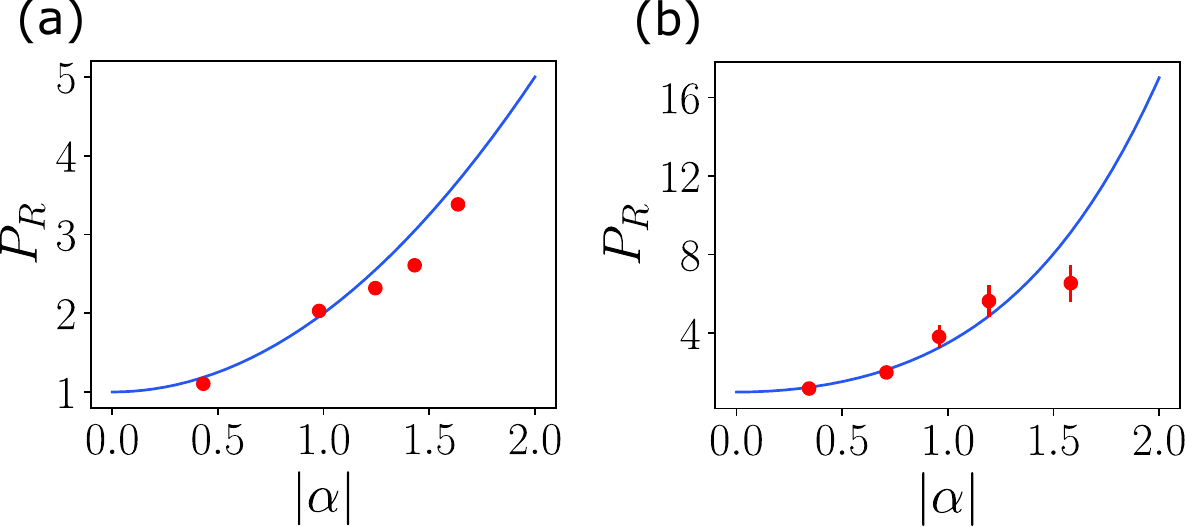}
  \caption{Relative heralding probability $P_R$ of $n$-photon-added coherent states is plotted for $n=1$ (a) and $n=2$ (b). Lines represent theoretical predictions and dots indicate experimental data. The experimental probability $P_R$ is determined from the coincidence rates of the heralding detectors while the input coherent state amplitude is estimated from homodyne measurements. The error bars represent one standard deviation. For some of the data, the error bars are smaller than the symbol size.}
  \label{fig:success_prob}
\end{figure}

\section*{Measurement of relative heralding probabilities}

Our estimation of  $|\alpha|$ and the calibration of the overall homodyne detector efficiency can be cross-checked by measurements of relative heralding probability 
\begin{equation}
P_R(\alpha,n)=\frac{P_H(\alpha,n)}{P_H(0,n)}=L_n(-|\alpha|^2).
\label{success_rate}
\end{equation}
This expression for $P_R$ follows from the formula (\ref{PH}) in the limit $\lambda \ll 1$.
The relative heralding probability  $P_R(\alpha,n)$ is an increasing function of $|\alpha|$, thanks to the effect of stimulated parametric down-conversion \cite{Linares2002}. 
In particular, for single-photon-added coherent state we obtain
\begin{equation}
P_R(\alpha,1)=1+|\alpha|^2,
\label{success_rate1}
\end{equation}
and for the two-photon-added coherent state we get
\begin{equation}
P_R(\alpha,2)=1+2|\alpha|^2+\frac{1}{2}|\alpha|^4.
\label{success_rate2}
\end{equation}
Note that the relative heralding  probability $P_R$ does not depend on the detection efficiency of the single-photon detectors in the heralding arm. The measurement of $P_R(\alpha,n)$ thus provides an independent absolute calibration of $|\alpha|$ that does not involve data from  homodyne detector. The $n$-photon-added coherent states were generated in several experimental runs and the reference heralding probability $P_H(0,n)$ was measured separately for each run and $n$.
The experimentally determined relative heralding probabilities of single- and two-photon-added coherent states are plotted in Fig. \ref{fig:success_prob} together with the theoretical predictions given by Eqs. (\ref{success_rate1}) and (\ref{success_rate2}). The good agreement between the theoretical predictions and the experimental results confirms the consistency of our estimation of  $\alpha$.

\begin{figure}
  \centering\includegraphics[width=0.75\linewidth]{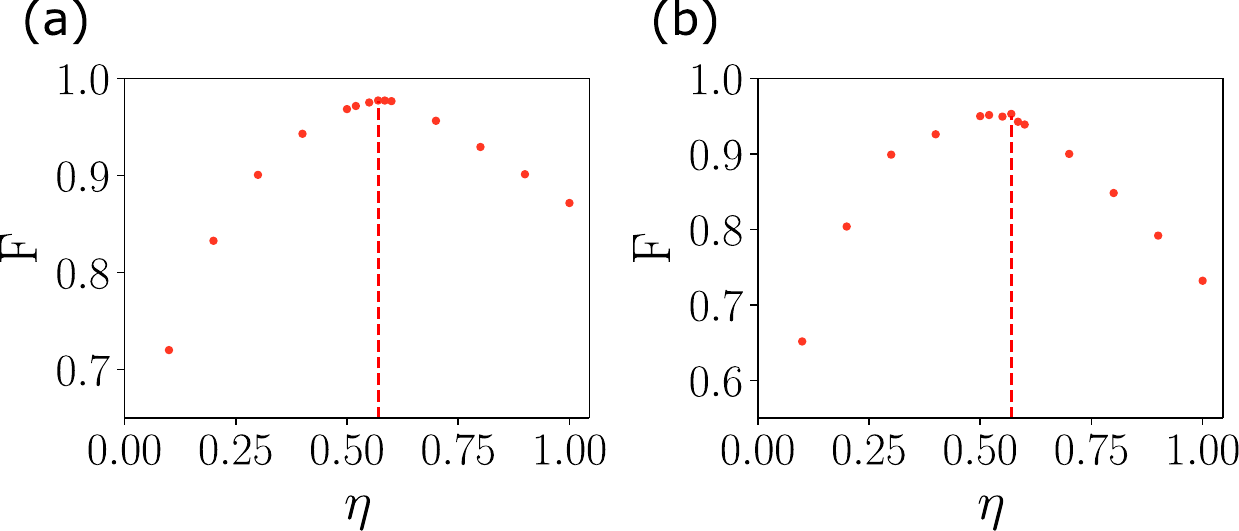}
\caption{Fidelities $F$ of the experimentally generated single-photon-added coherent state with $|\alpha|=1.43$ (a) and two-photon-added coherent state with $|\alpha|=1.20$ (b) are plotted as functions of the total effective homodyne detection efficiency $\eta$ assumed during quantum state reconstruction. In the reconstruction, losses amounting to $1-\eta$ are compensated for. The vertical red dashed lines indicate the fidelity maximum observed at $\eta=0.57$.}
\end{figure}

The estimation of $\alpha$ relies on correct calibration of the overall homodyne detection efficiency $\eta$. As discussed in detail in the Methods section of the main text, we estimate $\eta=0.57$, which corresponds to losses of $43\%$.  Consistency of our calibration of the homodyne detection efficiency $\eta$ can be further supported  by analysis of fidelity of the measured states with the theoretical states. Specifically, we can consider the dependence of the state fidelity on the amount of  losses compensated for during quantum state reconstruction. An example is provided in  Fig.~S3. When the assumed detection efficiency $\eta$ is decreased below $1$, the fidelity first  increases, because  we compensate for inefficient homodyne detection. The fidelity is maximized at  $\eta=0.57$, fully consistent with our total  loss estimate provided  in the Methods section.  When we reduce $\eta$ further, the fidelity drops down again, suggesting that we enter a regime where we overcompensate the true effective losses.

\section*{Additional results from quantum state reconstruction}

In the quantum state reconstruction we compensate for total detection losses of $43\%$ which includes $20\%$ losses imposed by the single-mode fiber inserted after the output of the nonlinear crystal. If we omit the effect of the fiber, the remaining detection losses read $29\%$, see the Methods section. For completeness, we provide here the values of state fidelities and purities of the generated $n$-photon-added coherent states obtained when we compensate only for losses of $29\%$ during the quantum state reconstruction. The results are summarized in the Table~SI below.

\begin{table}[h]
\caption{Fidelity $\tilde{F}$ and purity $\tilde{\mathcal{P}}$  of the experimentally generated $n$-photon-added coherent states with input amplitude $\tilde{\alpha}$ are listed. In contrast to the results reported in Table~I of the main text, only $29\%$ detection losses are compensated during the state reconstruction. Columns labeled $\Delta F=F-\tilde{F}$ and $\Delta \mathcal{P}=\mathcal{P}-\tilde{\mathcal{P}}$ specify the reduction of fidelity and purity with respect to the results reported in Table~I in the main text.}

\begin{ruledtabular}
\begin{tabular}{ccccccc}
$n$ & $|\tilde{\alpha}|$ & $\tilde{F}$ & $\Delta F$ & $\tilde{\mathcal{P}}$ & $\Delta \mathcal{P}$   \\
\hline
1 & 0.39(1) & 0.70(2) & 0.12 & 0.63(3) & 0.09   \\
1 & 0.88(1) & 0.84(2) & 0.06 & 0.83(3) & 0.03  \\
1 & 1.11(1) & 0.93(2) & 0.03 & 0.92(3) &  0.03 \\
1 & 1.28(2) & 0.95(1) & 0.02 & 0.96(3)  & 0.04  \\
1 & 1.47(1)& 0.92(2) & 0.02 & 0.91(5) &  0.01  \\
2 &  0.31(3) & 0.61(8) & 0.16 & 0.62(1)  & 0.28  \\
2 &  0.64(2) & 0.76(3) & 0.15 & 0.86(6) & 0.12  \\
2 &  0.86(2) & 0.83(3) & 0.11 & 0.89(5) & 0.08  \\
2 &  1.07(2) & 0.89(3) & 0.06 & 0.91(4) & 0.06  \\
2 &  1.42(3) & 0.88(3) & 0.03 & 0.88(5) & 0.02  \\
3 & 0.28(3) & 0.44(8) & 0.23 & 0.62(9) & 0.16 \\
3 & 0.85(3) & 0.78(6) & 0.08 & 0.78(8) & 0.09  \\
\end{tabular}
\end{ruledtabular}
\end{table}

For single-photon added coherent states, the state fidelities and purities are typically reduced only by a few percent while higher reductions occur for the two and three photon additions. 
Note that the estimated absolute values of coherent amplitudes  $|\tilde{\alpha}|$ are also reduced in comparison to the values reported in Table I of the main text, because we estimate $|\alpha|$ from homodyne measurements on the seed coherent states. Reduced loss compensation then implies lower estimates of $|\tilde{\alpha}|$. 
It can be seen from Table~SI that the reduction of fidelities and purities is strongest for states with small coherent amplitude $|\tilde{\alpha}|$. The observed behavior can be easily understood by noting that the fidelity of a Fock state $|n\rangle$ transmitted through a lossy channel with transmittance $\eta$ reads $\eta^n$, i.e. it decreases exponentially with increasing $n$. On the other hand, an ordinary coherent state after passing through a lossy channel is still a  pure coherent state.

\section*{Engineering coherent superpositions of additions and subtractions of $N$ photons}
Figure 7 in the main text illustrates a scheme for experimental realization of various coherent superpositions of different sequences of additions and subtractions of $N$ photons
$\hat{V}_j=\hat{a}^{N-j} \hat{a}^{\dagger N} \hat{a}^{j}$. For completeness,  we provide more details here,  following Ref. \cite{Fiurasek2009}. The operators $\hat{V}_j$ can be expressed as polynomials of $N$-th order in photon number operator,
\begin{equation}
\hat{V}_j =\prod_{k=1-j}^{N-j} (\hat{n}+k)=Q_j(\hat{n}).
\end{equation}
Note that $\hat{V}_j|m\rangle=0$ for $m<j.$ Consider now a target operation in the form of  polynomial  $S_N(\hat{n})=\sum_{k=0}^N s_k \hat{n}^k$.  To design this operation  as a superposition of $Q_j(\hat{n})$,
\begin{equation}
S(\hat{n})=\sum_{j=0}^N b_j Q_j(\hat{n}),
\end{equation}
one can proceed iteratively. Since $Q_j(0)=0$ for all $j>0$, we have $b_0=S(0)/Q_0(0)$. Next, consider $S(\hat{n})-b_0Q_0(\hat{n})$. This polynomial, when evaluated at $n=1$, yields the value of $b_1$, $b_1=[S(1)-b_0 Q_0(1)]/Q_1(1)$. More generally, we obtain the following iterative scheme,
\begin{equation}
b_j=\frac{S(j)-\sum_{k=0}^{j-1} b_kQ_k(j)}{Q_j(j)}.
\end{equation}
In the experiment, the polynomial will be further modulated by an exponentially decaying prefactor $\gamma^{\hat{n}}$, where $\gamma^2=(1-\lambda^2)T$ accounts for attenuation of the signal in the nonlinear crystal and at the photon-subtracting beam splitters with intensity transmittance $ T$ and reflectance $1-T$. Still, by suitably choosing the coefficients $b_j$ we can engineer an operation $\gamma^{\hat{n}}S(\hat{n})$ that will modulate the amplitudes of the first $N+1$ Fock states as desired.

As a specific example, consider implementation of the nonlinear sign gate on the subspace of Fock states $|0\rangle$, $ |1\rangle$ and $|2\rangle$, 
\begin{equation}
\hat{G}_{\mathrm{sign}} (c_0|0\rangle+c_1|1\rangle+c_{2}|2\rangle)= c_0|0\rangle+c_1|1\rangle-c_{2}|2\rangle.
\end{equation}
This gate, which is highly relevant in the context of optical quantum computing \cite{Knill2001},  leaves unchanged the states $|0\rangle$ and $ |1\rangle$ while it flips the sign of amplitude of the state $|2\rangle$. It is not difficult to check that the nonlinear sign gate is implemented by the following operator 
\begin{equation}
S_{\mathrm{sign}}(\hat{n})=1+\hat{n}-\hat{n}^2,
\end{equation}
which is a quadratic function of the photon number operator. 
The operation $S_{\mathrm{sign}}(\hat{n})$ can be expressed as combination of sequences of photon additions and subtractions as follows
\begin{equation}
S_{\mathrm{sign}}(\hat{n})= \frac{1}{2}\hat{a}^{ 2}\hat{a}^{\dagger 2}-\frac{1}{2}\hat{a}^{\dagger 2}\hat{a}^2-\hat{a}\hat{a}^{\dagger 2}\hat{a}.
\end{equation}

\medskip

\section*{Quantum statistical properties  of PACS}
Quantum statistical properties of the $n$-photon-added coherent states $|\alpha,n\rangle$ were comprehensively theoretically studied by  by Agarwal and Tara \cite{Agarwal1991}. Photon-added coherent states possess sub-Poissonian photon number distribution and can also exhibit  quadrature squeezing for specific values of $|\alpha|$.
For completeness,  we reproduce here the formulas for quadrature variances and photon number variance of the states $|\alpha,n\rangle$ from Ref. \cite{Agarwal1991}, together with a brief outline of their derivation. 

All the formulas presented below can be obtained with the help of the analytical expression for normally ordered moments of annihilation operators, 
\cite{Perina1984},
\begin{equation}
\langle \alpha|\hat{a}^m\hat{a}^{\dagger n}|\alpha\rangle=\alpha^{m-n} n! L_n^{(m-n)}(-|\alpha|^2),
\label{antinormal}
\end{equation}
which is valid for $m\geq n$. Here
\begin{equation}
L_n ^{(k)}(x)=\sum_{j=0}^n {n+k \choose n-j} \frac{(-x)^j}{j!}
\end{equation}
is the generalized Laguerre polynomial. In particular, it follows that the  normalization factor $\mathcal{N}_n(\alpha)=[ \langle \alpha|\hat{a}^n\hat{a}^{\dagger n}|\alpha\rangle]^{-1/2}$ 
can be expressed as \cite{Agarwal1991}
\begin{equation}
\mathcal{N}_n(\alpha) =\left[n!L_n(-|\alpha|^2) \right]^{-1/2},
\end{equation}
where $L_n(x)$ denotes the ordinary Laguerre polynomial of degree $n$. With the help of the expression (\ref{antinormal}) we can also derive  a compact and elegant analytical expression for the noiseless amplification gain
\begin{equation}
g_n(\alpha)=\frac{\langle \alpha,n|\hat{a}|\alpha,n\rangle}{\alpha}.
\end{equation}
Since $\langle \alpha, n|\hat{a}|\alpha,n\rangle=\mathcal{N}_n^2(\alpha) \langle \alpha|\hat{a}^{n+1}\hat{a}^{\dagger n}|\alpha\rangle$,  we obtain 
\begin{equation}
g_n(\alpha)=\frac{L_n^{(1)}(-|\alpha|^2)}{L_n(-|\alpha|^2)}.
\label{gnformulageneral}
\end{equation}

\subsection*{Quadrature variances}
 For a state with complex amplitude  $\alpha=|\alpha|e^{i\theta}$ we can define the  amplitude quadrature operator $\hat{x}$ and the phase quadrature operator $\hat{p}$ as 
\begin{equation}
\hat{x}=\hat{a}^\dagger e^{i\theta}+\hat{a}e^{-i\theta}, \qquad \hat{p}=i(\hat{a}^\dagger e^{i\theta}-\hat{a}e^{-i\theta}).
\end{equation}
This choice guarantees that $\langle \hat{x}\rangle=2|\langle \hat{a}\rangle|$ and $\langle \hat{p}\rangle=0$.
It follows that the variances of $\hat{x}$ and $\hat{p}$ can be expressed as
\begin{equation}
\langle (\Delta \hat{x})^2\rangle= 2\langle \hat{a}\hat{a}^\dagger\rangle-1+ e^{-2i\theta}\langle \hat{a}^2\rangle+e^{2i\theta}\langle \hat{a}^{\dagger2}\rangle-4|\langle \hat{a}\rangle|^2, \qquad 
\langle (\Delta\hat{p})^2\rangle= 2\langle \hat{a}\hat{a}^\dagger\rangle-1- e^{-2i\theta}\langle \hat{a}^2\rangle -e^{2i\theta}\langle \hat{a}^{\dagger 2}\rangle.
\label{varxpdefinition}
\end{equation}
For PACS, the required moments of creation and annihilation operators can be determined from Eq. (\ref{antinormal}),
\begin{equation}
\langle \hat{a}\rangle= \alpha \frac{L_n^{(1)}(-|\alpha|^2)}{L_n(-|\alpha|^2)}, \qquad \langle \hat{a}^2\rangle= \alpha^2 \frac{L_n^{(2)}(-|\alpha|^2)}{L_n(-|\alpha|^2)}, \qquad
 \langle \hat{a}\hat{a}^\dagger\rangle= (n+1)  \frac{L_{n+1}(-|\alpha|^2)}{L_n(-|\alpha|^2)}.
\label{momentsV}
\end{equation}
If we combine Eqs. (\ref{varxpdefinition}) and (\ref{momentsV}), we obtain the following expressions for the quadrature variances \cite{Agarwal1991},
\begin{equation}
V_x=\frac{2}{L_n^{(0)}(-|\alpha|^2)}\left[ |\alpha|^2 L_n^{(2)}(-|\alpha|^2)+(n+1)L_{n+1}(-|\alpha|^2)   \right]-1-4|\alpha|^2 \left[\frac{L_n^{(1)}(-|\alpha|^2)}{L_n(-|\alpha|^2)}\right]^2
\end{equation}
and
\begin{equation}
V_p=\frac{2}{L_n(-|\alpha|^2)}\left[ (n+1)L_{n+1}(-|\alpha|^2) -|\alpha|^2 L_n^{(2)}(-|\alpha|^2)   \right]-1.
\end{equation}
With the help of formulas (\ref{momentsV}) one can also easily verify that the covariance of $\hat{x}$ and $\hat{p}$ vanishes, 
\begin{equation}
\langle \Delta\hat{x}\Delta\hat{p}+\Delta\hat{p}\Delta\hat{x}\rangle=0.
\end{equation}

\subsection*{Fano factor}

The photon number distribution of the photon-added coherent states can be characterized by the  Fano factor
\begin{equation}
F_N=\frac{\langle( \Delta \hat{n})^2\rangle}{\langle \hat{n}\rangle},
\label{fano}
\end{equation}
where $\langle( \Delta \hat{n})^2\rangle$ is the variance of the mean photon number. If $F<1$ then the state exhibits sub-Poissonian photon number statistics. Explicit formula for Fano factor of the $n$-photon-added coherent states reads \cite{Agarwal1991}
\begin{equation}
F_N=\frac{(n+2)(n+1)L_{n+2}(-|\alpha|^2)-2L_n(-|\alpha|^2)}{(n+1)L_{n+1}(-|\alpha|^2)-L_n(-|\alpha|^2)}-(n+1)\frac{L_{n+1}(-|\alpha|^2)}{L_n(-|\alpha|^2)}-2.
\label{Fanothpacs}
\end{equation}
The first and second moments of the photon number operator required for calculation of the Fano factor can be straightforwardly obtained from the following expressions for the antinormally ordered moments of $\hat{a}$ and $\hat{a}^\dagger$ \cite{Agarwal1991}
\begin{eqnarray}
&\displaystyle{\langle \hat{a}\hat{a}^\dagger\rangle=\langle \hat{n}+1\rangle=(n+1)\frac{L_{n+1}(-|\alpha|^2)}{L_n(-|\alpha|^2)},} & \nonumber \\
&\displaystyle{ \langle \hat{a}^2\hat{a}^{\dagger 2}\rangle=\langle (\hat{n}+1)(\hat{n}+2)\rangle=\langle \hat{n}^2+3\hat{n}+2\rangle=(n+2)(n+1)\frac{L_{n+2}(-|\alpha|^2)}{L_n(-|\alpha|^2)}.} &
\end{eqnarray}
 All the experimentally generated $n$-photon-added coherent states listed in Table I of the main text exhibit sub-Possonian photon number distribution  with the Fano factor ranging from $0.23$ to $0.92$. The experimentally observed Fano factors are generally larger than the theoretical prediction (\ref{Fanothpacs}) due to various experimental imperfections.

\section*{Stellar rank witness based on fidelity with single-photon-added coherent state}
Here we derive analytical expression for the maximum fidelity of the single-photon-added coherent state $\hat{a}^\dagger |\alpha\rangle$ with a Gaussian state. First recall that up to coherent displacements, the single-photon-added coherent state is equivalent to coherent superposition of vacuum and single-photon states,
\begin{equation}
|\psi\rangle=\frac{1}{\sqrt{1+|\alpha|^2}}\left(\alpha^\ast|0\rangle+|1\rangle\right).
\label{psi}
\end{equation}
Maximum fidelity of the state (\ref{psi}) with a Gaussian state has been derived numerically in Ref. \cite{Chabaud2021}. Here we provide analytical treatment. The most general pure Gaussian state is a squeezed coherent state, that can be expanded  in the Fock basis as follows,
\begin{equation}
|\phi\rangle_G=\sum_{n=0}^\infty\frac{1}{ \sqrt{ 2^n n! \cosh r}} (e^{i\theta}\tanh r)^{n/2} \exp\left( -\frac{|\beta|^2}{2} +\frac{e^{-i\theta}}{2} \beta^2 \tanh r\right) 
H_n\left(\frac{e^{-i\theta/2}\beta}{\sqrt{\sinh(2r)}}\right) |n\rangle.
\label{phiG}
\end{equation}
Here $\beta \in \mathbb{C}$ is complex amplitude, $r\in \mathbb{R}$ is real squeezing constant, $\theta$ is the squeezing angle, and  $H_n(x)$ denotes Hermite polynomial.
The overlap between states (\ref{psi}) and (\ref{phiG}) reads
\begin{equation}
\langle \psi|\phi\rangle_G=\frac{1}{\sqrt{(1+|\alpha|^2)\cosh r}} \exp\left( -\frac{|\beta|^2}{2} +\frac{e^{-i\theta}}{2} \beta^2 \tanh r\right)  \left[ \alpha+\frac{\beta}{\cosh r}\right].
\end{equation}
Without loss of generality, we can assume that $\alpha$ is real and positive, and denote $a=|\alpha|$. The fidelity 
\begin{equation}
F=|\langle \psi|\phi\rangle_G|^2
\end{equation}
is then maximized if $\beta$ is also real and positive and $\theta=0$. We are thus left with the function
\begin{equation}
F= \frac{1}{(1+a^2)\cosh^3 r} (a\cosh r+b)^2\exp\left[-(1-\tanh r)b^2\right]
\end{equation}
that needs to be optimized with respect to parameters $r$ and $b=|\beta|$.
The extremality conditions
\begin{equation}
\frac{\partial F}{\partial r}=0, \qquad \frac{\partial F}{\partial b}=0, 
\end{equation}
yield the following two equations for the optimal parameters $b$ and $r$,
\begin{equation}
\left(ab+\frac{b^2}{\cosh r}\right)(\tanh r-1)+\frac{1}{\cosh r}=0, \qquad
\frac{b^2}{\cosh^2 r}\left(a+\frac{b}{\cosh r}\right) -\tanh r \left (a+ \frac{3b}{\cosh r}\right)=0.
\end{equation}
If we combine these two equations, we can express $b$ in terms of $r$ and $a$,
\begin{equation}
b=a \cosh r \frac{1-e^{-2r}}{3e^{-2r}-1},
\end{equation}
and the squeezing constant $r$ can be found by solving a cubic equation
\begin{equation}
a^2z^3+9z^2-(6+a^2)z+1=0,
\end{equation}
where $z=e^{-2r}$. This equation has three real roots,  two positive and one negative. The two positive roots both lie in the interval $(0,1)$. In order to find the maximum fidelity $F$ one has to consider both positive roots and choose the root that maximizes the fidelity.

\end{document}